\documentclass[manuscript,screen]{acmart}

\AtBeginDocument{%
  \providecommand\BibTeX{{%
    \normalfont B\kern-0.5em{\scshape i\kern-0.25em b}\kern-0.8em\TeX}}}

\setcopyright{acmlicensed}
\copyrightyear{2024}
\acmYear{2024}
\acmDOI{XXXXXXX.XXXXXXX}

\usepackage{multirow}
\usepackage{threeparttable}
\usepackage{float}
\usepackage{graphicx}
\begin{document}

\title{NARRepair: Non-Autoregressive Code Generation Model for Automatic Program Repair}


\author{Zhenyu Yang}
\affiliation{%
  \institution{Shandong University}
  \city{Qingdao}
  \country{China}}
\email{yangzycs@mail.sdu.edu.cn}

\author{Zhen Yang}
\affiliation{%
  \institution{Shandong University}
  \city{Qingdao}
  \country{China}}
\email{zhenyang@sdu.edu.cn}

\author{Zhongxing Yu}
\affiliation{%
  \institution{Shandong University}
  \city{Qingdao}
  \country{China}}
\email{zhongxing.yu@sdu.edu.cn}


\begin{abstract}
Recent years have witnessed a surge of research efforts on Automatic Program Repair(APR), which promises to reduce software development costs and improve software reliability. With the advancement of deep learning techniques, the performance of APR techniques has reached a new level. Previous deep learning-based APR techniques essentially used a sequence-to-sequence model to modify program sentences in the Autoregressive(AR) manner, which predicts future values based on past values. Due to the manner of word-by-word generation, the AR-based APR technique has a huge time delay and thus cannot fix bugs in real time. This negative consequence overshadows the widespread adoption of APR techniques in real-life software development. 

To address the issue, we aim to apply the Non-Autoregressive(NAR) method to the APR task, which can output target code in a parallel manner to avoid huge inference delays. However, the naive use of the NAR manner for the APR task suffers from the issue of compromised patch quality. To effectively adapt the NAR manner for the APR task, we in this paper propose NARRepair, the first customized NAR code generation model for the APR task. The NARRepair features three major novelties. First, NARRepair is guided by repair actions to alleviate the issue of over-correction, i.e., correct code can instead be modified into wrong ones. Second, to alleviate the issue of lacking inter-word dependency information associated with NAR manner, NARRepair extracts this dependency information on top of the Abstract Syntax Tree (AST) for generating words in parallel while maintaining the correctness of the code syntax. Finally, to alleviate the issue of lacking contextual information, NARRrepair obtains contextual information about words through two-stage decoding for improving the accuracy of the patch generation process. 

We evaluated NARRepair on three widely used datasets in the APR community, including Defects4J v1.2, Defects4J v2.0, and QuixBugs. The results show that 1)  compared to AR-based APR techniques, the inference speed of NARRepair has been increased by 5.4-15.1 times in the CPU environment and 6.2-18.6 times in the GPU environment, and 2) NARRepair has fixed 69, 41, and 23 bugs for Defect4J v1.2, Defect4J v2.0, and QuixBugs respectively, which are respectively 90\%, 95\%, and 82\% of the optimal model. Overall, the results show that our technique can significantly improve the inference speed while maintaining high repair accuracy.
\end{abstract}





\maketitle

\section{Introduction}
Program defects are inevitable during the software development process, and 
recent years have witnessed a surge of research efforts on Automatic Program Repair(APR) to alleviate this issue \cite{le2011genprog, nguyen2013semfix, mechtaev2015directfix, kim2013automatic, liu2018mining, le2016history, koyuncu2019ifixr, long2017automatic, XinASE, Elixir,Prophet, rsrepair, overfitting3, overfitting4,wencontext,jiang2018shaping,angelix,gaoissta,10172854,YuEmSE,yutse,yubot}. Automatic program repair aims to automatically change the buggy code into correct code and promises to reduce software development costs and improve software reliability. To achieve this, different mechanisms have been explored in the ARP area, notably including search-based repair \cite{le2011genprog,yuan2018arja,jobstmann2005program}, constraint-based repair \cite{nguyen2013semfix,mechtaev2015directfix,xuan2016nopol}, and template-based repair \cite{kim2013automatic,koyuncu2020fixminer,liu2018mining}. Enlightened by the huge success of deep learning in a wide variety of application areas, researchers have also investigated the use of deep learning for the APR task in the past few years \cite{yu2021deeprepair, chen2019sequencer, zhu2021syntax, CODIT, lutellier2020coconut} and impressive repair results have been obtained. 
In particular, given that sequence-to-sequence neural network models have powerful learning capabilities, a majority of these Deep Learning(DL)-based APR techniques basically are built on top of sequence-to-sequence models. For example, SequenceR \cite{chen2019sequencer} is a program repair method based on a sequence-to-sequence structure with a copy mechanism and Rewardrepair \cite{ye2022neural} uses program compilation and test execution information to help models repair buggy code text. 

The use of sequence-to-sequence models suggests that these Deep Learning(DL)-based APR techniques essentially modify program code in the 
Autoregressive(AR) manner, which predicts future values based on past values. In other words, AR generative models generate words one by one in sequence. This use of AR manner leads to the inability of real-time repair and huge time delays for repairing real-life complex bugs, which typically involves modifications to long code sequences. These two negative consequences overshadow the widespread adoption of Deep Learning(DL)-based APR techniques in real-life software development. For instances, in the field of embedded systems (e.g., FPGA and Robots), several researchers have emphasized the importance of real-time program  repair\cite{gaimon1989real,steinbauer2005real, nazar2015improving}. For achieving the goal, they however only proposed solutions on top of domain specificities and did not fundamentally solve the issue of insufficient real-time repair performance. Therefore, the field of program repair still lacks a general solution to the issue of real-time program repair.

\begin{figure}[]
\centering
\includegraphics[width=0.8\linewidth]{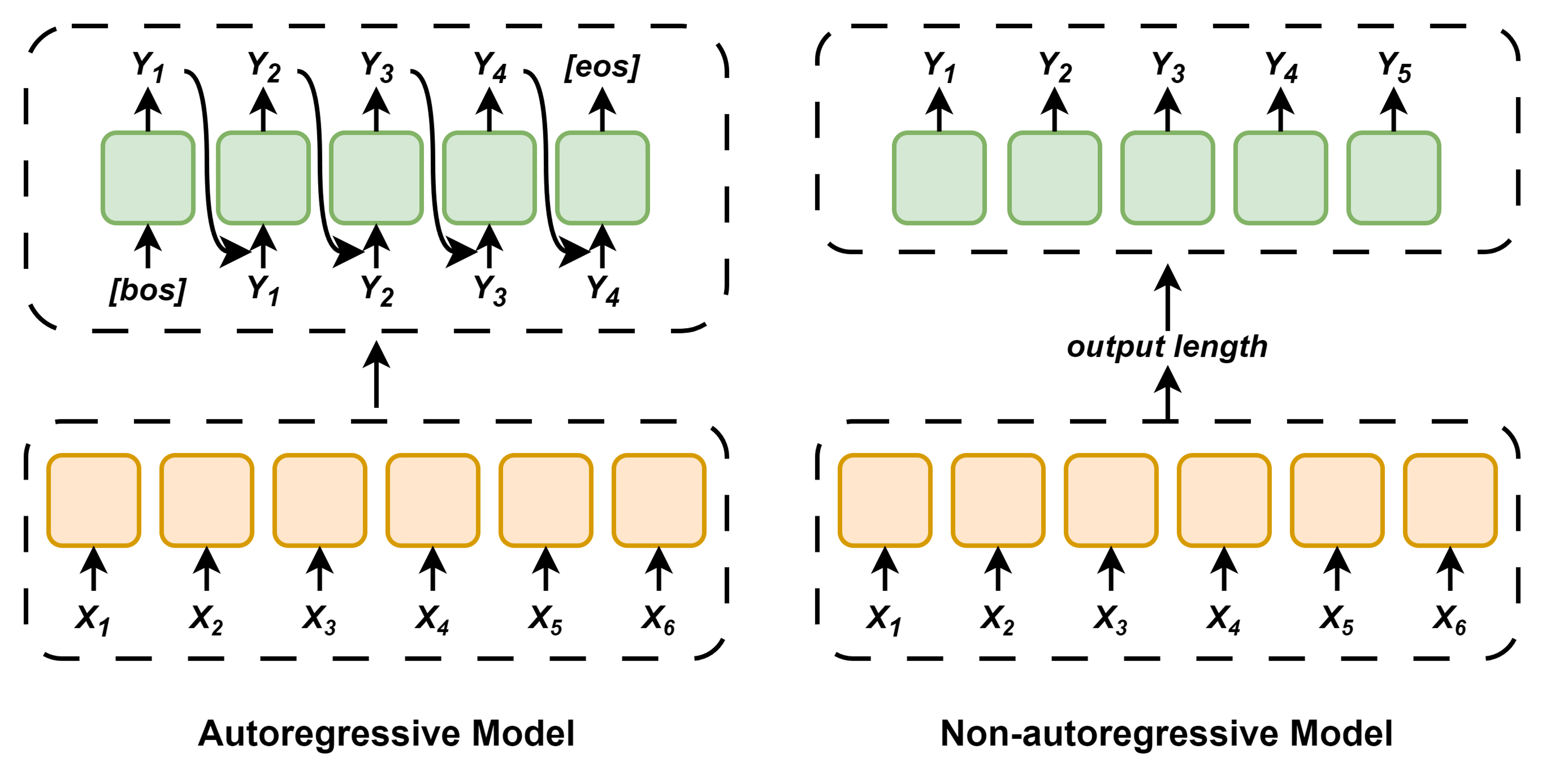}
\caption{\label{fig:frog1}The inference process of autoregressive (AR) model and non-autoregressive (NAR) model.}
\end{figure}

To solve this problem, inspired by the work on Non-Autoregressive Machine Translation(NAT)\cite{gu2018non, li2019hint}, we propose to design techniques on top of the Non-Atoregressive(NAR) model. Compared to the AR model, the NAR model \cite{gu2018non} generates words of target text in parallel and thus can greatly improve the speed of model inference. The inference process of the AR model and the NAR model is illustrated in Figure \ref{fig:frog1}. While the NAR models have been widely used in the field of machine translation, naive use of NAR models (in the field of machine translation) for the APR task regrettably results in poor performance according to our experiments. In particular, three major limitations of NAR models for the APR task are as follows.

\textbf{Limitation 1: Over-correction problem.} The process of machine translation involves converting text in one language into another language, so the NAR model of machine translation will modify each word in the source text into a word in other languages. However, unlike machine translation, the words that need modification account for only a small portion of the buggy code in the APR task. It is difficult for the NAR model to identify all the correct words in the buggy code and keep them without making any modifications. Consequently, the NAR model will suffer from the over-correction problem, that is, many correct words can instead be modified into wrong ones in the generated target code. This is one of the main reasons why NAR models perform poorly in the APR task. 

\textbf{Limitation 2: Unable to obtain inter-word dependency information.} Unlike the AR model which generates words sequentially, the NAR model generates words in parallel. In other words, the NAR model does not know what words are elsewhere in the sentence when it generates a word. Thus, the NAR model improves the inference speed at the cost of losing the inter-word dependency information. However, the inter-word dependency information is vital for sentence grammar correctness. Therefore, how to ensure the grammatical correctness of the code generated by the NAR model is an urgent problem to be solved \cite{saharia2020non}. Given that programs feature formal grammar and rich structural dependency, losing inter-word dependency information will significantly deteriorate the performance of the APR task. 

\textbf{Limitation 3: Unable to obtain contextual information of the buggy text.} The effectiveness of the DL-based APR techniques considerably relies on using the contextual information to modify wrong words. However, since the NAR model generates words in parallel and processes context words and wrong words in the same way, it cannot effectively obtain the contextual information. Hence, the model is unable to make use of the contextual information of the code when modifying the wrong words, leading to compromised performance. 

To overcome the above limitations of the NAR model for the APR task, we in this paper propose a novel NAR model, named NARRepair, for generating fixed code sequences quickly while ensuring accuracy. Similar to most NAR models, NARRepair uses the encoder-decoder structure on top of the transformer model. However, NARRepair features three major novelties for addressing the above three limitations in order.

\textbf{Novelty 1: Action-guided repair} (corresponding to limitation 1). The main reason why the NAR model modifies the correct word into the wrong word is that the word probability distribution learned by the model is unstable. When generating a word in the target text, the NAR model generates the probabilities of all words in the dictionary and selects the one with the highest probability. In other words, the model can correctly retain the word only when the probability of the original word is the highest. However, a dictionary that contains tens of thousands of words will inevitably have noise, and thus the model does not know whether a specific word should be retained or not. To address this problem, we propose an action-guided repair method. We divide the repair processes into four actions: ``keep'', ``replace'', ``insert'', and ``delete''. Instead of directly predicting the correct word, NARRepair predicts the repair action corresponding to each word in the buggy code. That is, our model uses predicted actions to guide the decoder to generate words. This method noticeably reduces the complexity of the probability distribution that the model needs to learn. In particular, the dimension is reduced from the number of words in the dictionary to the number of actions. Consequently, the learning burden of the model is significantly reduced, and the issue of wrongly modifying the correct words can be effectively alleviated. 

\textbf{Novelty 2: Inter-word dependency building based on abstract syntax tree} (corresponding to limitation 2). In abstract syntax tree (AST), words corresponding to certain AST nodes are connected to other words through parent AST nodes. We assume that a node in the AST has an association with its parent node(s) and has the strongest association with the nearest parent node. Furthermore, the nearest common parent node of a word pair has a strong association with both words and can represent the dependency information between the two words. To enable the NAR model to obtain dependency information between words, especially dependency information at the semantic level, we propose an inter-word dependency building method based on AST. We use the nearest common parent nodes to represent the dependency between words and add dependency information into the inference process. Therefore, NARRepair can obtain information about other words while generating words in parallel. This method makes the target text generated by NARRepair smoother and more grammatical. 

\textbf{Novelty 3: Two-stage decoding} (corresponding to limitation 3). To overcome the problem that the NAR model cannot obtain the contextual information of the buggy text, we propose a two-stage decoding method. This method divides the decoding process of the model into two stages. The first stage of the decoding produces fixed code text of poor quality. We keep the words with high confidence and mask the remaining words. Words with high confidence include (1) multiple adjacent words for which repair actions are ``keep'' and (2) words whose prediction probabilities are greater than the threshold. Masked language models \cite{devlin2018bert,feng2020codebert,liu2019roberta} have been proven effective in multiple works to obtain the contextual information of masked words. Therefore, our method can obtain the contextual information about words with low confidence. NARRepair uses the contextual information to regenerate words to improve the repair accuracy in the second decoding stage. 


To evaluate the performance of NARRrepair, we assess its inference speed and accuracy on three widely used datasets for the APR task: Defect4J v1.2 \cite{just2014defects4j} containing 395 bugs, Defect4J v2.0 containing 420 bugs, and QuixBugs \cite{lin2017quixbugs} containing 40 bugs. In terms of inference speed, compared with other autoregressive APR models, the inference speed of NARRepair is 5.4-15.1 times faster and 6.2-18.6 times faster in CPU and GPU environments respectively. In terms of accuracy, NARRepair fixes 69 bugs from the Defect4J v1.2 dataset, which represents 90\% of the optimal model (Tare); NARRepair fixes 45 bugs from the Defect4J v2.0 dataset, which represents 95\% of the optimal model (Rewardrepair); NARRepair fixes 23 bugs from the QuixBugs dataset, which represents 82\% of the optimal model (AlphaRepair). In addition, with regard to accuracy, the repair results of the NARRepair model for the three datasets are better than other advanced NAR models. These experimental results demonstrate that the proposed NARRepair model maintains high accuracy while greatly increasing the inference speed. 

In summary, our contributions to this work are as follows:
\begin{itemize}
    \item We propose the NARRepair model for the APR task. To the best of our knowledge, NARRepair is the first NAR model designed for the APR task. 
    \item We propose three techniques to overcome the limitations of the naive use of NAR model for the APR task, including 1) the repair action predictor for alleviating the over-correction problem, 2) the inter-word dependency extractor for alleviating the issue of lacking inter-word dependency information, and 3) the two-stage decoder for alleviating the issue of lacking contextual information.
    \item We evaluate the performance of the NARRepair model on three widely used datasets for the APR task. The results show that NARRrepair can greatly reduce inference latency while maintaining high accuracy.
\end{itemize}

\section{Related Work}
\textbf{APR and Deep Learning-based APR.} Given the time-consuming and error-prone nature of program debugging \cite{debugging,multiple-fault}, APR techniques have been proposed to reduce software development costs and improve software reliability. Recent years have witnessed a surge of APR techniques rooted in different disciplines, notably including search-based repair, constraint-based repair, and template-based repair. Search-based APR techniques \cite{le2011genprog,yuan2018arja,jobstmann2005program, jobstmann2005program}, also known as heuristic-based techniques, treat generating patches as finding feasible solutions in a predefined search space. Constraint-based APR techniques \cite{xuan2016nopol,nguyen2013semfix,wei2010automated,pei2011code,mechtaev2015directfix} guide the repair process by first developing a set of constraints and then solving these constraints to derive the patches. Template-based APR techniques \cite{kim2013automatic,liu2018mining,liu2019avatar,koyuncu2019ifixr,koyuncu2020fixminer,le2016history,liu2019tbar,long2015staged} rely on various targeted repair templates (also called ``fix pattern'' in the literature) to generate patches, which have good repair effects on specific types of software defects.  

Enlightened by the huge success of deep learning in a wide variety of application areas, researchers have also investigated the use of deep learning for the APR task in recent years. As a result, there are an abundance of DL-based APR techniques in the literature. Gupta et al. \cite{gupta2017deepfix} propose DeepFix, an APR model to repair compilation defects in C language code. DeepFix is a multi-layered sequence-to-sequence neural network that directly predicts defect locations and the corresponding correct code for the defect. White et al. \cite{yu2021deeprepair} propose an APR model named DeepRepair, which infers code similarity through deep learning. This technology can sort code fragments based on their similarities with suspicious elements and can convert statements by mapping identifiers outside the scope to similar identifiers within the scope. Chen et al. \cite{chen2019sequencer} propose a technology named SequenceR for end-to-end APR on top of the sequence-to-sequence model, which uses abstract context to simulate the process of analyzing and fixing bugs conducted by developers. Lutellier et al. \cite{lutellier2020coconut} propose CoCoNuT, a technology for APR using a neural machine translation model based on convolutional neural networks. Zhu et al. \cite{zhu2021syntax} propose Recoder, which constrains the output of the APR model via syntax rules to repair fine-grained erroneous sentences. Ye et al. \cite{ye2022neural} propose RewardRepair, which adds test information to the model to ensure that candidate patches are compilable. Xia et al.\cite{xia2022less} treat program repair tasks as text fill-in-the-blanks and generate patches based on contextual information. Zhu et al.\cite{zhu2023tare} propose a type-aware model for the APR task to reduce unusable patches.

At present, a majority of these DL-based APR techniques basically are built on top of sequence-to-sequence models and generate correct code in the AR manner. The AR model requires that the output of each step waits for the output of the previous position in order, resulting in slow reasoning. Consequently, this use of AR manner leads to the inability of real-time repair and huge time delays for repairing real-life complex bugs, which typically involves modifications to long code snippets. These negative consequences create obstacles to the adoption of DL-based APR techniques in real-life software development and maintenance.

\vspace{1.0mm}
\noindent
\textbf{Non-autoregressive Models.} The purpose of NAR models is to reduce inference time by generating target sentences in parallel. Gu et al. \cite{gu2018non} propose the concept and the first NAR model, which assumes that all words in the target sentence are independent of each other and can output all target words in parallel in one step. Bao et al. \cite{bao2019non} propose PNAR, which uses latent variables to capture the arrangement information of target words for making the arrangement of target words more appropriate. Shu et al. \cite{shu2020latent} use a spherical Gaussian to generate latent variables for each input word to increase the dependence between words in the target sentence. Ran et al. \cite{ran2021guiding} use latent variables to establish the position information of the target word. Ma et al. \cite{ma2019flowseq} use generative flow to model latent variables containing rich information. Stern et al. \cite{stern2019insertion} propose a NAR model based on insertion operations, which generates a subsequence of the final result sequence through iteration at each step until all insertion operations are empty and the iteration ends. However, given the three major limitations outlined in Section 1, naively using existing NAR models for the APR task cannot obtain satisfactory results. Therefore, we propose the NARRepair model in this paper to meet the unique needs of the APR task.
\begin{figure*}[]
\centering
\includegraphics[width=1\textwidth]{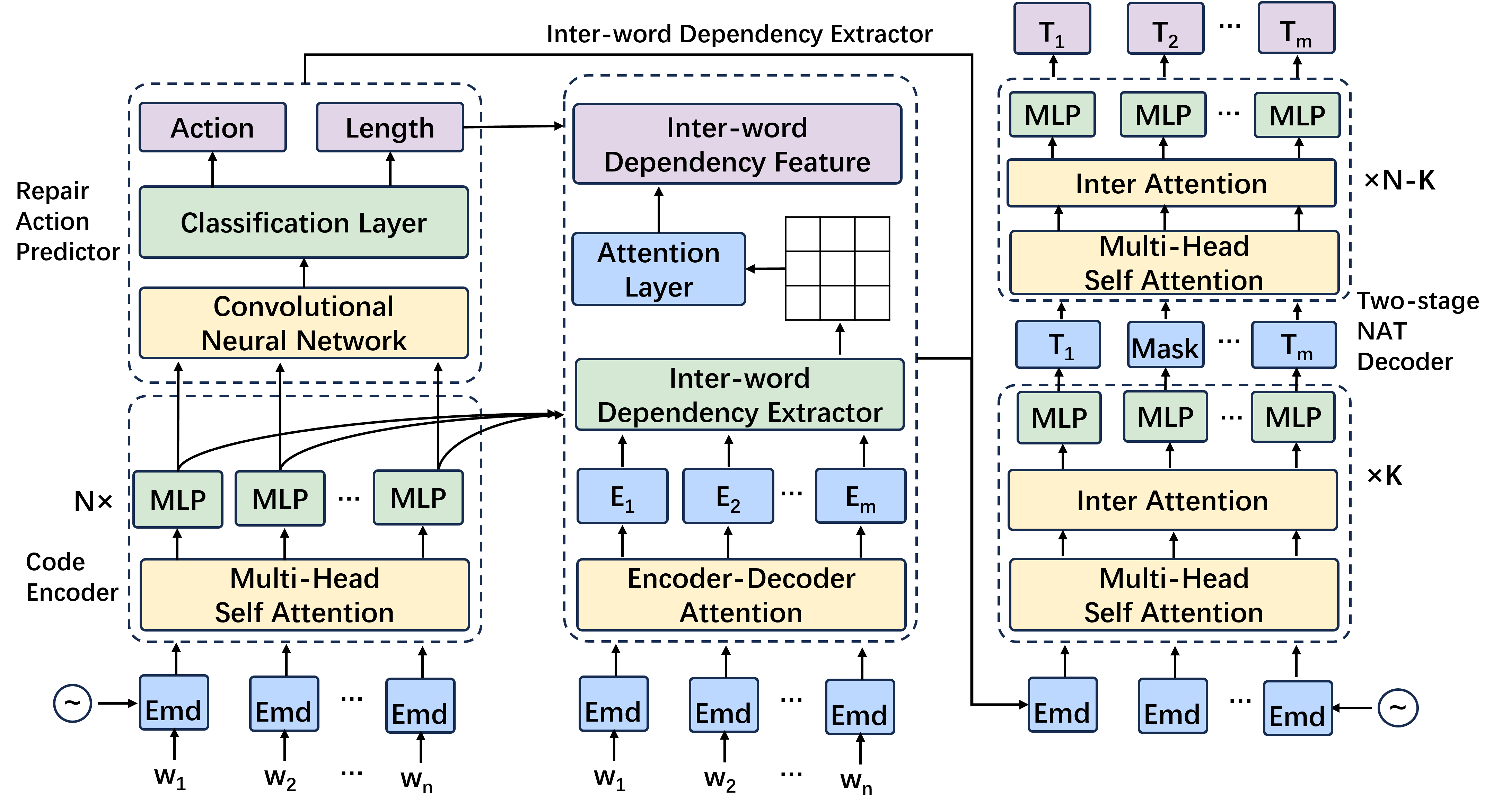}
\caption{\label{fig:frog2}An overview of the NARRepair architecture. }
\end{figure*}
\section{NARRepair}
In this section, we will introduce the NARRepair model, which uses NAR to generate code text in parallel to improve the inference speed. The model mainly consists of four parts: code encoder, repair action predictor, inter-word dependency extractor, and two-stage NAR decoder. Figure \ref{fig:frog2} shows the structure of the NARRepair model. The process of NARRepair is as follows:
\begin{itemize}
    \item Code encoder embeds the buggy code into feature vectors (§3.1).
    \item Given the buggy code feature, the repair action predictor predicts repair action and output length for each word (§3.2).
    \item According to the output length, the inter-word dependency extractor first generates the feature vector of the repaired code text. Then, the extractor obtains the inter-word dependency information and fuses it with the word feature vector to obtain the word feature vector with dependency information (§3.3).
    \item Given the word feature vector from the previous step, the two-stage decoder generates all repaired words (§3.4).
\end{itemize}
In the following sections, we will elaborate on the structure of the NARRepair.
\begin{figure}
\centering
\includegraphics[width=0.5\linewidth]{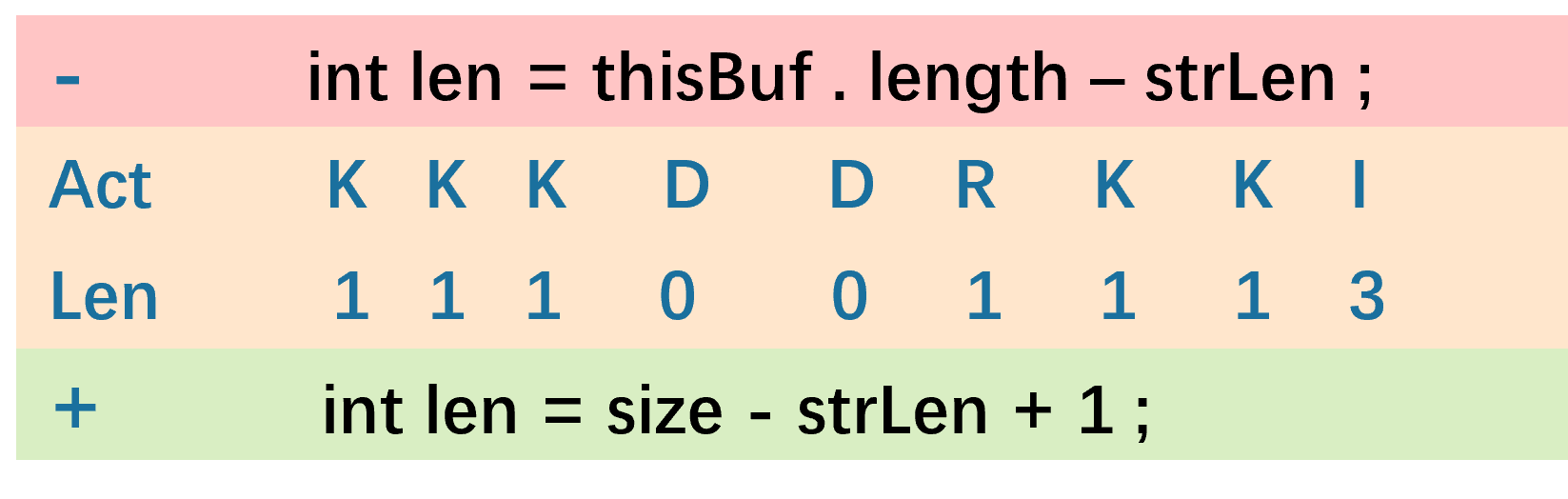}
\caption{\label{fig:frog3} A bug (Lang 61 from Defects4J) fixed by NARRepair with repair action predictor. }
\end{figure}
\subsection{Code Encoder}
The code encoder can extract features from the buggy code text ${W}_{i:n}$ and convert ${W}_{i:n}$ into a word embedding ${E}_{i:n}$. The output word embedding ${E}_{i}$ can be used
for the prediction and tagging of subsequent modules. We use the encoder part of the transformer model \cite{vaswani2017attention} as the code encoder of the model.

Here, we briefly give an overview of the encoder of the transformer model. The encoder of the transformer is composed of multiple identical layers, and each layer has two sub-layers. The first is a multi-head self-attention layer that fuses word features by calculating the attention weight between word feature vectors. The second sub-layer is a feedforward neural network used to normalize the output of the model. Residual links are used between sub-layers. The operation process of the code encoder can be defined as
\begin{equation}\label{eq1}
{E}_{i:n}=Encoder({W}_{i:n}+{W}_{pos})
\end{equation}
and the operation of each layer of the encoder can be expressed as
\begin{align}\label{eq2}
{X}^{l}_{attention}&={X}^{l-1}_{hidden}+Attention({X}^{l-1}_{hidden})\\
{X}^{l}_{hidden}&=Feedforward({X}^{l}_{attention})\\
{X}^{0}_{hidden}&={W}_{i:n}+{W}_{pos}
\end{align}
where ${X}_{pos}$ is position embedding, $Attention$ is self-attention layer, and $Feedforward$ is feedforward neural network layer.

\subsection{Repair Action Predictor}
The repair action predictor can predict the repair action for each word in the buggy code text. The content of the repair action is divided into two parts: the type of repair action and the repair length. We classify all repair actions into 4 categories: ``keep'', ``insert'', ``delete'', and ``replace''. The repair length represents the number of generated repair words corresponding to each fixed word. Typically, the repair length for actions ``replace'' and ``keep'' is 1; the repair length for action ``delete'' is 0; the repair length for action ``insert'' is the number of words inserted.
Figure \ref{fig:frog3} gives an example of repair action prediction. As shown in Figure \ref{fig:frog3}, the repair action predictor predicts the repair action and repair length for "Lang-61" buggy code in the Defect4j dataset. Compared with NAR models in machine translation that need to predict the probabilities of all words in the dictionary, the repair action predictor only needs to predict the probabilities of four repair actions. When the predicted action is ``keep'', the model does not need to change the words. This method effectively alleviates the problem of modifying the correct words into the wrong ones. 

Regarding the model structure, since convolutional neural networks have the advantage of effectively acquiring local features, the repair action predictor uses a convolutional neural network \cite{kalchbrenner2014convolutional} to extract the word features after receiving the output of the encoder. Then, the classification layer predicts the repair action and the repair length for each word in the buggy code text separately. The detailed operations are as follows:
\begin{align}\label{eq2}
{X}_{feature}&=ConV({E}_{1:n})\\
{Act}_{1:n}={Linear}&_{1}(Dropout(Relu({X}_{feature})))\\
{Len}_{1:n}={Linear}&_{2}(Dropout(Relu({X}_{feature})))
\end{align}
where $ConV$ is the convolutional neural network layer, ${Linear}_{1}$ is a fully connected layer whose output dimension is the number of repair actions, and ${Linear}_{2}$ is a fully connected layer whose output dimension is the maximum length. We set the maximum prediction length to be 10. We use the cross-entropy method to compute the loss between the predictor output and the ground truth as:\\
\begin{align}\label{eq2}
{L}_{lenth}&=-\sum ^{n}_{i} {log{p}_{lenth}({l}_{i}|X,\, M)}\\
{L}_{act}&=-\sum ^{n}_{i} {log{p}_{act}({a}_{i}|X,\, N)}
\end{align}
where $M$ and $N$ are model parameters, $X$ is the input word feature vector, ${l}_{i}$ is the repair length of the i-th word, and ${a}_{i}$ is the repair action of the i-th word.
\begin{figure*}
\centering
\includegraphics[width=1\textwidth]{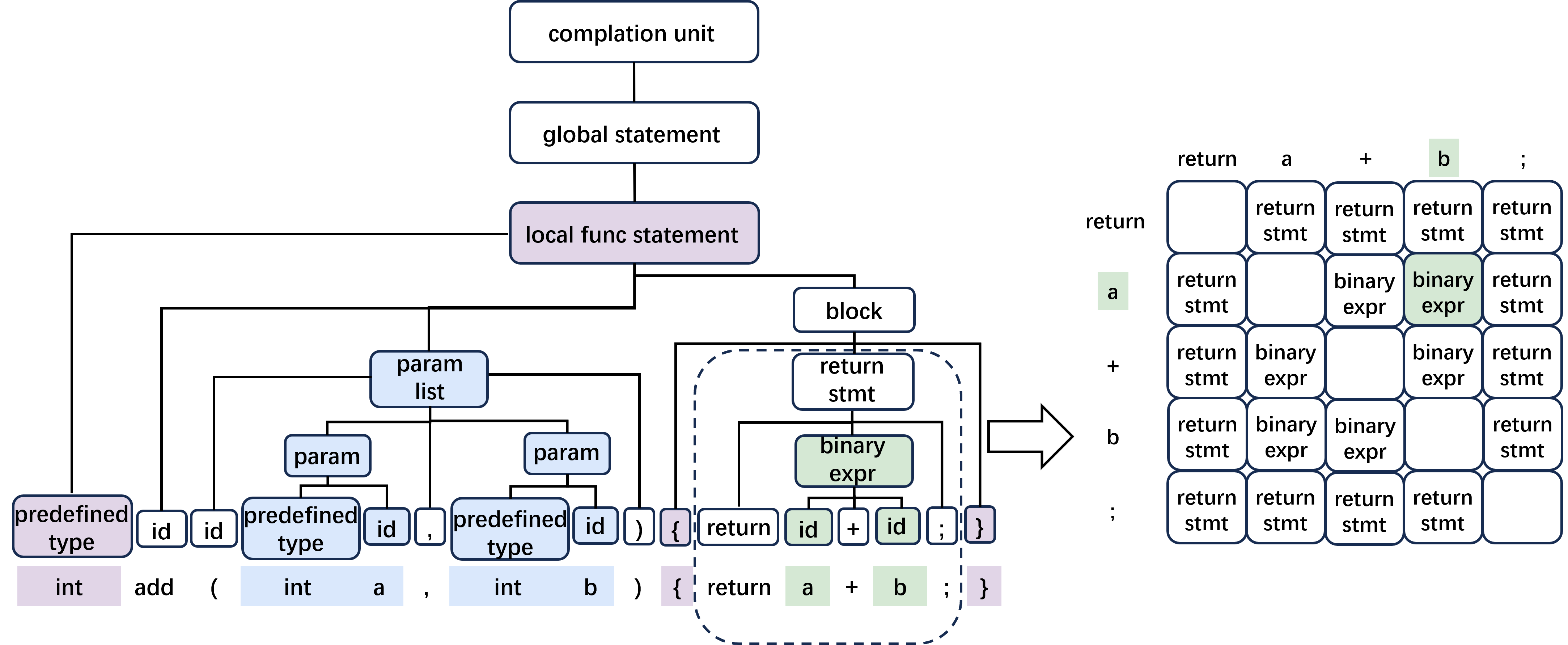}
\caption{\label{fig:frog4}The example of generating AST and inter-word dependency matrix for the text ``int add(int a, int b) \{return a+b;\}''.}
\end{figure*}
\subsection{Inter-word Dependency Extractor}    
The inter-word dependency extractor can learn the dependency information between word pairs through the nearest common parent nodes in the AST. To get the nearest parent node of word pairs, we need to generate the AST for the code text. Given the code text, we use the program analysis tool Tree-sitter\cite{tree-sitter} to extract its AST ${T}_{1:m}$. Figure \ref{fig:frog4} shows the code text ``int add(int a, int b) \{return a+b;\}'' and its corresponding AST. In this example, the nearest common parent of ``int a'' and ``int b'' is ``param list''; the nearest common parent of ``a'' and ``b'' on the right is ``binary expr''. We list the nearest common parent nodes of all word pairs and generate a word dependency matrix. We show the inter-word dependency matrix of ``return a+b'' in the right part of Figure \ref{fig:frog4}. The relationship between a certain word ``n'' and another word ``m'' is equivalent to that between ``m'' and ``n'', so the dependency matrix is symmetric. We use the obtained dependency matrix as the ground truth to train the inter-word dependency extractor. 
\begin{figure}
\centering
\includegraphics[width=0.5\linewidth]{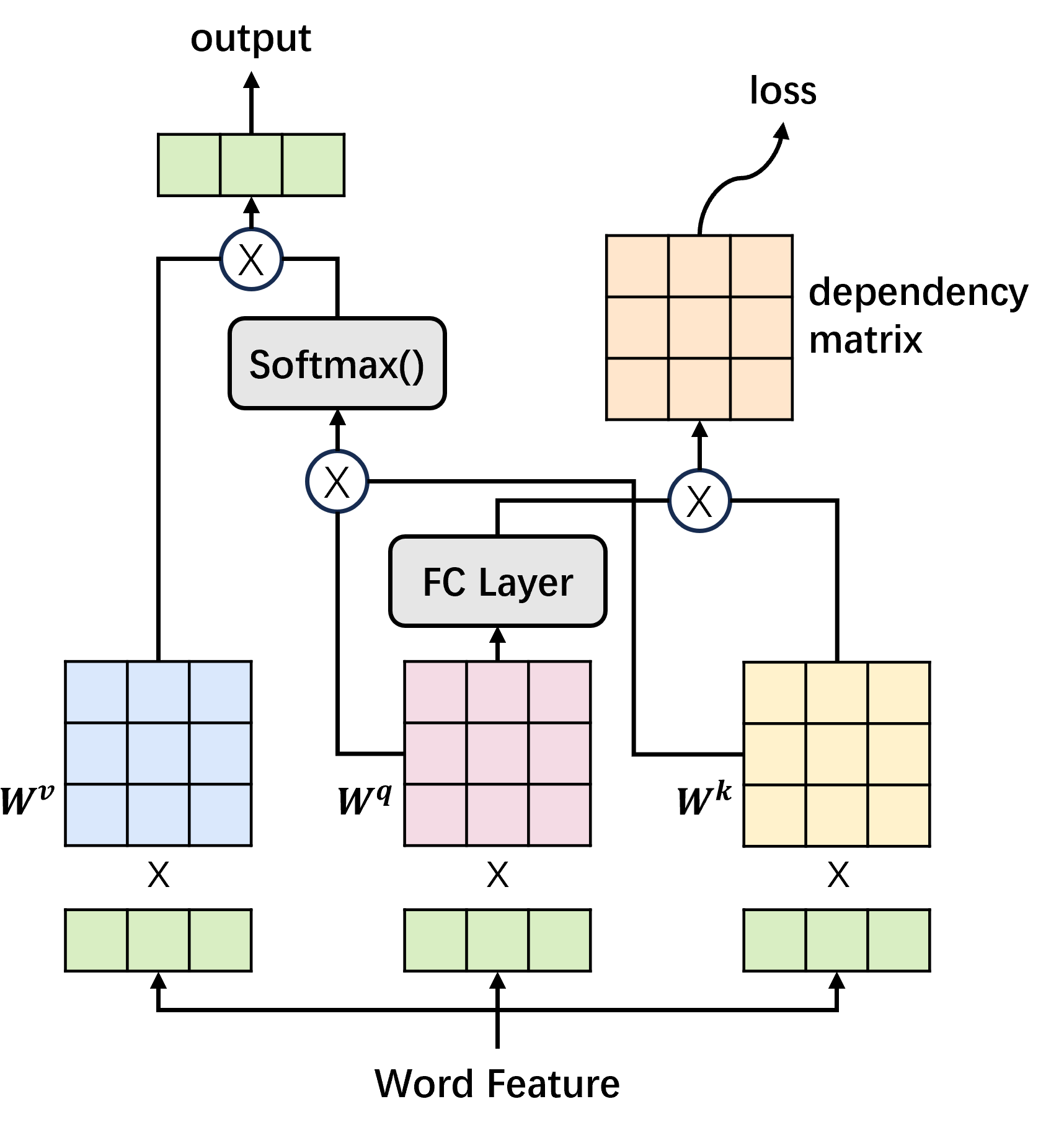}
\caption{\label{fig:frog5}Structure of the Inter-word Dependency Extractor.}
\end{figure}
Given the feature vector ${X}_{1:n}$ of the faulty text, we first use the encoder-decoder attention module to obtain the feature vector ${D}_{1:m}$ of the target text. Then, we use the inter-word dependency extractor to predict the nearest common parent nodes of word pairs as dependency information. Inspired by the work of Dozat et al. \cite{dozat2016deep}, the inter-word dependency extractor uses an attention mechanism to model dependency information. For each target text feature, the extractor maps the feature using the query ${W}^{q}$ and the key ${W}^{k}$. The dot product of maps Q and K predicts the word dependency matrix. To facilitate training, we construct an index table of parent nodes in the abstract syntax tree and replace the parent nodes in the matrix with index values as ground truth. Figure \ref{fig:frog5} shows the structure of the inter-word dependency extractor. The detailed operations are as follows:
\begin{align}\label{eq2}
Q&={W}^{q}*{D}_{1:m}\\
K&={W}^{k}*{D}_{1:m}\\
M_{dependency}&=Linear3(Q)*K^{T}
\end{align}
where ${W}_{q}$ and ${W}_{k}$ are the weight matrix, and $Linear3$ is a fully connected layer used to convert the feature dimension into the dimension of the number of parent nodes in the AST. After obtaining the dependency information, we use an attention machine to fuse it into word features in order for the decoder to obtain the dependency information about other words when generating words. The detailed operations are as follows:
\begin{align}\label{eq2}
score &= softmax(\frac {Q{K}^{T}} {\sqrt {d}})\\
V&={W}^{v}*{D}_{1:m}\\
{H}_{1:m} &= \sum ^{n}_{i=1} {score*V}
\end{align}
where ${W}^{v}$ is the weight matrix, $softmax$ is the normalization function for calculating fractions. We compute the loss of the inter-word dependency extractor as:\\
\begin{align}\label{eq2}
{L}_{depend}=-log(M|D,E)
\end{align}
where D is the input feature vector, E is the parameter of the extractor, and M is the output word dependency matrix.

\subsection{Two-stage NAR Decoder}
The two-stage NAR decoder decomposes the normal NAR decoder into two parts for step-by-step decoding. 
The purpose of the first stage of decoding is to generate a preliminary result and retain words with high confidence. For the text generated by the first stage of decoding, we retain two parts of the words whose confidence levels are high: (1) multiple adjacent words for which repair actions are ``keep'' and (2) words whose prediction probabilities are greater than a threshold.
\begin{figure}
\centering
\includegraphics[width=0.40\linewidth]{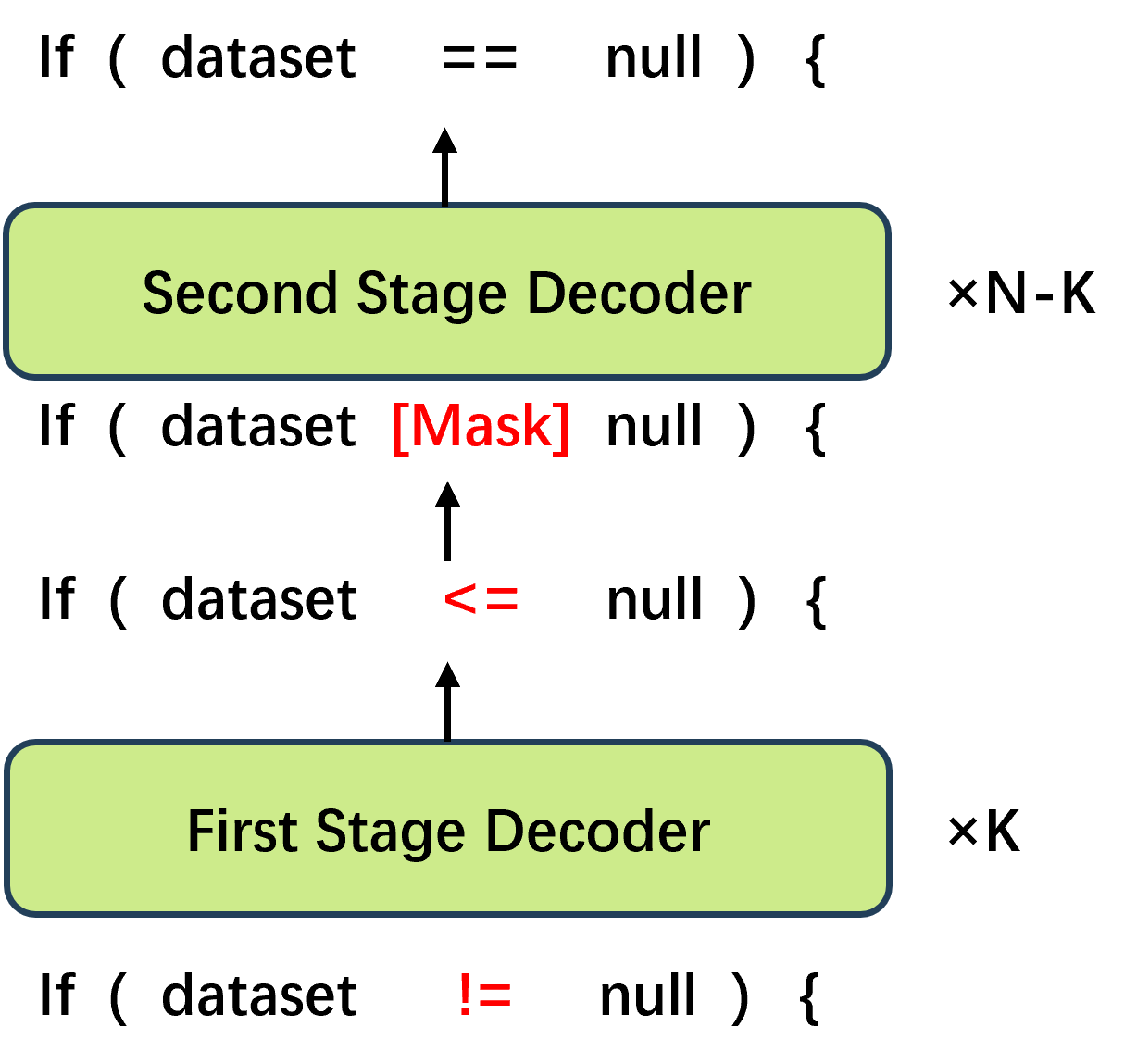}
\caption{\label{fig:frog6}Decoding Process of the Two-stage NAR Decoder}
\end{figure}
We set the threshold to be 0.7 in our experimental evaluation. We retain these high-confidence words and use them as context for the remaining low-confidence words. Since the remaining words may be wrong, the purpose of the second decoding is to regenerate the words with low confidence based on the contextual information. We mask the remaining words with the [Mask] tag. Literature about mask language models \cite{devlin2018bert,feng2020codebert,liu2019roberta,guo2020graphcodebert} has proven that the attention mechanism can obtain contextual information for words with the [Mask] tag. Hence, the result of the second stage of decoding based on contextual information will be more accurate. 

Take the code text ``if (dataset !=null) \{'' as an example, Figure \ref{fig:frog6} shows the decoding process of the two-stage NAR decoder. First, the first stage decoder generates the preliminary result ``if (dataset <= null) \{'' and considers ``<='' as having low confidence. Then, the second stage decoder masks the ``<='' with the ``[Mask]'' tag to obtain the context information and generate the correct result.
Assuming that the given input feature is ${H}_{1:m}$, the specific operations of the first stage decoder are as follows:
\begin{align}\label{eq2}
{E}_{first}={Decoder}_{first}&({H}_{1:m})={Layer}^{1:n-k}_{decoder}({H}_{1:m})\\
{P}_{first}&=Softmax({E}_{first})
\end{align}
where ${Layer}^{1: n-k}_{decoder}$ is the 1st to $n-k$th layer of the decoder, ${E}_{first}$ is the feature vector outputted by the first stage of decoding, and ${P}_{first}$ is the probability distribution of the first stage of decoding. 
We further mask the result of the first stage of decoding and input it into the second stage decoder as follows:
\begin{align}\label{eq2}
{E}_{mask}&= MaskFunc({E}_{first})\\
{E}_{second}={Decoder}_{second}&({E}_{mask})={Layer}^{n-k:n}_{decoder}({E}_{mask})\\
{P}_{second}&=Softmax({E}_{second})
\end{align}
where $MaskFunc$ is the masking function used to replace words with low confidence with the [MASK] tag, ${Layer}^{n-k : n}_{decoder}$ is the $n-k$th to $n$th layer of the decoder, ${E}_{second}$ is the feature vector outputted by the second stage of decoding, and ${P}_{second}$ is the word probability distribution generated by the second stage of decoding. 
We compute the loss based on the output of the decoder as:
\begin{align}\label{eq2}
{{L}_{dec}}_{}&=-\sum ^{m}_{i} {log({P}_{dec}(R|H,\partial ))}
\end{align}
where $R$ is the output result of the decoder, $H$ is the hidden feature, and $\partial$ is the decoder parameters.

\subsection{Training and Inference}
During the training process, the three sub-modules of the NARRepair model are jointly learned. The final loss function of NARRepair is as follows:
\begin{align}\label{eq2}
L={L}_{dec}+\alpha({L}_{act}+{L}_{length})+\lambda{L}_{depend}
\end{align}
where $\alpha$ and $\lambda$ are the hyperparameters used to adjust the importance of each training loss. We set $\alpha$ to be 0.1 and $\lambda$ to be 0.1.

During inference, the repair action predictor firstly predicts the repair action and output length for each word in the buggy code. Then, based on the output length, the encoder-decoder attention layer transforms the buggy code's feature vector into the fixed code's feature vector. Next, the inter-word dependency extractor generates fixed code feature vectors with inter-word dependency information through AST. Finally, the two-stage decoder outputs the fixed code based on the repair action and the fixed code feature vector with inter-word dependency information. For each buggy code, we generate 200 patches and test whether they contain the correct patches.
 
\section{Evaluation Setup}
To prove the validity of our idea, we train the NARRepair model and evaluate its performance. In this section, we introduce the evaluation setup.

\subsection{Research Questions}
To begin with, we give the research questions explored in this paper.   

\vspace{1.0mm}
\noindent
\textbf{RQ1: Compared with other DL-based APR models and NAR models, what is the performance of NARRepair for the APR task?} This is an overall question about the performance of the NARRepair model. For this question, we evaluate the NARRepair model on three widely used datasets for APR tasks and compare its performance with that of other DL-based APR models and NAR models.

\vspace{1.0mm}
\noindent
\textbf{RQ2: How does each module contribute to the final result of NARRepair?} For this question, we gradually remove submodules from the complete NARRepair model and investigate the contribution of each submodule.

\vspace{1.0mm}
\noindent
\textbf{RQ3: How does NARREpair help with existing pre-trained models?}
Pre-trained models \cite{feng2020codebert, yupre} and large language models \cite{touvron2023llama} have had a significant impact on the field of software engineering. 
This question wants to investigate whether our idea can help the pre-training model to improve the inference speed alike. Since the large language model is highly similar to the pre-trained model, the results obtained for pre-trained models can be generalized to large language models to some extent. For this question, we first load the weights of multiple mainstream pre-trained models for programming languages into the NARRepair model and train the model on the dataset, and then compare the result of a single pre-trained model with that of the NARRepair model with loaded parameters.

\vspace{1.0mm}
\noindent
\textbf{RQ4: Can the NARRepair model effectively alleviate the over-correction problem?} One of the important limitations we target is the over-correction problem. To investigate the effectiveness of our idea for alleviating this limitation in detail, we compare the number of correct words that NARRepair changed into incorrect words before and after removing the repair action predictor module and the two-stage decoder module. 

\vspace{1.0mm}
\noindent
\textbf{RQ5: Whether the nearest parent node is closely related to its child node in the AST?} We made the hypothesis that the nearest parent node has a close relationship with its child nodes. This question aims to investigate the validity of this hypothesis. 

\subsection{Data Collection}
We use the dataset published by selfAPR\cite{ye2022selfapr} on GitHub as our model dataset, which contains code text pairs (buggy and correct) generated for common types of program errors by self-supervision. We discard some data for which we could not generate the AST and retain 837, 059 pieces of data. Since this dataset is machine-generated, we collect additional datasets from real projects on GitHub to supplement this dataset. To prevent data leakage, we remove projects related to Defect4J\cite{just2014defects4j} and QuixBugs\cite{lin2017quixbugs}. Finally, our dataset contains 1.52 million instances, and we use 90\% of the dataset as training data and 10\% as validation data.

We use three widely used datasets for the APR task to evaluate the model performance. The first one is Defect4J v1.2 \cite{just2014defects4j}, which contains 395 bugs in real Java projects. The second one is Defect4J v2.0 \cite{just2014defects4j}, which contains 438 additional bugs compared to Defect4J v1.2. The third one is QuixBugs \cite{lin2017quixbugs}, which contains 40 Java bugs and is introduced specifically to test the robustness of the model on data distributions besides Defect4J bugs.

\subsection{Data Preprocessing}
For training data, we need to generate repair action(s) and inter-word dependency matrix for each piece of data to facilitate model learning. For generating repair actions, we use dynamic programming to calculate the edit distance between the buggy code text and the fixed code text. After calculating the edit distance, we use backtracking to output the specific repair action for each step. For example, for a code pair containing the buggy code text ``if ( result != null )'' and the fixed code text ``if ( ! result . isNotype ( ) )'', we can accordingly get the required repair actions on top of this procedure: Keep, Keep, Insert, Replace, Replace, Insert. For the inter-word dependency matrix, we first use the Tree-sitter tool \cite{tree-sitter} to generate an AST for each correct code piece and obtain the path from each word to the root node. Then, we compare the paths of different words and add the nearest common parent of the word pair to the dependency matrix. We establish the repair action(s) and inter-word dependency matrix for each piece of training data and feed the established data to the model. 

\subsection{Knowledge Distillation}
Due to the conditional independence assumption of the NAR model, it is difficult to capture the multimodal distribution of target text, which is called the ``multi-modality problem'' in the literature\cite{gu2018non}. For example, the buggy code ``Node block = NodeUtil.getFunctionBody ( fnNode ) ;'' corresponds to two fixed code pieces ``Node argsNode = NodeUtil.getFnParameters ( fnNode ) ;'' and ``Node block = fnNode.getLastChild ( ) ;''. Since the NAR model outputs results in a parallel manner and lacks information about other locations, for the above example, the NAR model may output ``argsNode'' in the second location and ``fnNode.getLastChild'' in the fourth location. While these outputs all correspond to the correct code, they do not combine correctly. This situation is very common in the dataset and seriously affects the performance of the NAR model.

To alleviate this problem, we refer to previous NAR works \cite{gu2018non,wang2019non,qian2020glancing} and use the knowledge distillation method to process the training dataset. First, we train the CodeT5-large pre-trained model with the training dataset. Then, we use the text generated by the trained CodeT5-large on the original training dataset as the distilled training dataset. This kind of processing can remove the noise in the dataset as much as possible, and only retain the most correctly fixed code. The model trained on the distilled training dataset can learn the knowledge of program repair more accurately. To account for the model's robustness, we train the NARRepair model on the distilled training dataset and the original training dataset simultaneously and learn primarily from the distilled training dataset.

\subsection{Baselines}
We select mainstream models in the APR and NAR fields as the baselines for our experiments. In the field of APR, we select 6 models that are often used as baselines: SequenceR \cite{chen2019sequencer}, CoConut \cite{lutellier2020coconut}, Rewardrepair \cite{ye2022neural}, Recoder \cite{zhu2021syntax}, AlphaRepair \cite{xia2022less}, and Tare\cite{zhu2023tare}. These baseline models include models with simple structure and fast inference speed like SequenceR, and models with complex structure, good performance but slow inference speed like AlphaRepair. In addition, we also select Tare and AlphaRepair, which (to our knowledge) are the state-of-the-art models on the Defect4J and QuixBugs datasets respectively. Following previous work\cite{ye2022neural,xia2022less,zhu2023tare}, we select only comparing models with similar parameter sizes and exclude large models with an extremely large amount of parameters (such as ChatGPT\cite{chatgpt} and LLAMA\cite{touvron2023llama}) in the baselines. For the repair accuracy of the baselines, we directly adopt the results published by the corresponding papers to facilitate fair comparison. For example, since SequenceR only publishes the experimental results for the Defect4j v1.2 data set under the perfect defect location assumption, we directly adopt this result and consider the results for other datasets and scenarios as unknown. For the inference speed of the baselines, we re-run the models provided by the authors to establish the needed time. 

In the field of NAR, we have selected five advanced models proposed in recent years: DePA \cite{zhan2022non}, OAXE-NAR \cite{du2021order}, Fully-NAR \cite{gu2020fully}, SNAR \cite{liu2021enriching}, and CTC \cite{saharia2020non}. To ensure the fairness of the experiment, following other APR works \cite{ye2022neural,xia2022less}, we load the weights of the pre-trained model CodeT5 \cite{wang2021codet5} into NARRepair and other NAR baseline models. Our purpose is to respond to what we mentioned in the introduction that the naive use of NAR models (in the field of machine translation) for the APR task is not suitable and proves the advancement of our model. We implemented five NAR baseline models to evaluate their performance and inference speed on the APR task.

\subsection{Metrics}
To verify the correctness of the generated patches, we follow the previous work of the APR community to verify patches using two methods: test cases and manual analysis. A patch is considered valid if it successfully passes all test cases. A patch is considered correct if it successfully passes manual analysis. The latter is a subset of the former. We enumerate the specific numbers of the two types of patches in the experimental results.

\subsection{Implementation Details}
We use Pytorch \cite{pytorch} to implement the NARRepair model. During training, same as in previous work, we use the Adam optimizer \cite{kingma2014adam} to update the model parameters. During the optimization process, to feed the model as much data as possible, we set the batch size and epoch in all our experiments to be 50 and 200 respectively. As the training process proceeds, the learning rate is adjusted (ranging from 0 to 0.00005) to adapt to the learning situation at different stages of the model. The maximum sequence length is set to be 512, and the word out of range is ignored. After experimental verification, we set the maximum repair length to be 20.

\section{Evaluation Results}
In this section, we present the main experimental results. First, we give the comparison results about the accuracy and inference latency of the NARRepair model and the baseline models. Next, we introduce ablation experiments to show the effectiveness of each module of the model. Then, we show that the NARRepair model can help with existing pre-trained models in improving the inference speed. After that, we demonstrate the effectiveness of NARRepair model for alleviating the issue of over-correction generally associated with the NAR models. Finally, we verify the existence of a close relationship between parent nodes and child nodes in the AST. In terms of equipment, we use 4 NVIDIA RTX 3090 GPUs and the CPU is ``Intel(R) Xeon(R) Gold 6226R CPU @ 2.90GHz''.

\begin{table}[]
\renewcommand{\arraystretch}{1.5}
\centering
\caption{\label{tab:t1}Number of Corrected Bugs and Inference Speed for Different Models.}
\resizebox{\textwidth}{!}{
\begin{tabular}{cccccccccc}
\hline
\multicolumn{2}{c}{\multirow{3}{*}{Model}} & \multicolumn{6}{c}{Datasets}                                                        & \multicolumn{2}{c}{Latency on TestSet}                  \\ \cline{3-10} 
\multicolumn{2}{c}{}                       & \multicolumn{3}{c}{Perfect FL}           & \multicolumn{3}{c}{without Perfect FL}   & \multirow{2}{*}{CPU}             & \multirow{2}{*}{GPU} \\ \cline{3-8}
\multicolumn{2}{c}{}                       & Defect4J v1.2 & Defect4J v2.0 & QuixBugs & Defect4J v1.2 & Defect4J v2.0 & QuixBugs &                                  &                      \\ \hline
\multirow{5}{*}{NAT Mdoel}  & ReorderNAT   & 18/30         & 14/21         & 8/10     & 17/29         & 9/17          & 6/8      & 17.9x                            & 18.6x                \\ \cline{2-10} 
                            & SNAT         & 20/37         & 18/26         & 8/10     & 17/33         & 11/19         & 6/9      & 21.3x                            & 22.6x                \\ \cline{2-10} 
                            & Fully-NAT    & 35/61         & 22/37         & 10/13    & 25/41         & 15/26         & 9/10     & 15.2x                            & 16.5x                \\ \cline{2-10} 
                            & OAXE-NAT     & 27/56         & 17/38         & 9/11     & 19/36         & 12/24         & 8/10     & 14.4x                            & 15.3x                \\ \cline{2-10} 
                            & DePA          & 31/60         & 20/38         & 11/15    & 23/39         & 14/26         & 9/10     & 13.8x                            & 15.1x                \\ \hline
\multirow{6}{*}{APR Model}  & SequenceR    & 14/19         & -             & -        & -             & -             & -        & 1.0x                             & 1.0x                 \\ \cline{2-10} 
                            & CoCoNut      & 44/85         & -             & 13/20    & -             & -             & -        & -1.1x(5.9x)                      & -1.2x(7.4x)          \\ \cline{2-10} 
                            & Rewardrepair & 44/-          & 43/-          & 20/-     & 27/-          & 24/-          & -        & -1.4x(7.6x)                      & -1.5x(9.3x)          \\ \cline{2-10} 
                            & Recoder      & 64/-          & -             & -        & 49/90         & 19/46         & 17/17    & -2.2x(11.8x)                     & -2.6x(16.1x)         \\ \cline{2-10} 
                            & AlphaRepair  & 74/109        & 36/50         & 28/30    & 50/90         & -             & -        & \multicolumn{1}{l}{-2.8x(15.1x)} & -3.0(18.6x)          \\ \cline{2-10} 
                            & Tare         & 77/-          & -             & -        & 62/115        & 32/84         & 27/27    & \multicolumn{1}{l}{-2.4x(13.0x)} & -2.7x(16.8x)         \\ \hline
our                         & NARRepair    & 69/79         & 41/48         & 23/25    & 51/68         & 26/43         & 20/22    & 5.4x                             & 6.2x                 \\ \hline
\end{tabular}
}
\begin{tablenotes}
        \footnotesize
        \item[*] In the cells, (1) x/y: x denotes the number of correct patches, and y denotes the number of plausible patches that pass all test cases, (2) a(b): a denotes the difference in inference speed with regard to the fastest autoregressive model SequenceR, b denotes the difference in inference speed with regard to the NARRepair model, positive numbers mean speeding up and negative numbers mean slowing down, (3) the ‘-’ symbol indicates that the number has not been published in the corresponding article.  
\end{tablenotes}
\end{table}
\subsection{(RQ1) Results of Comparison with the Baselines}

\noindent
\textbf{Results with Perfect Fault Localization.}
We first compare the repair performance of the NARRepair model with that of the baseline models under the perfect fault localization scenario. The results are presented in Table \ref{tab:t1}. The NARRrepair model correctly repairs 69, 41, and 23 buggy programs for the three datasets Defect4J v1.2, Defect4J v2.0, and QuixBugs respectively. Compared with the NAR models in the field of machine translation, the NARRepair model outperforms the optimal NAR model by 197\%, 186\%, and 209\% in the number of repaired programs for the three datasets respectively. Compared with other DL-based APR techniques that use the autoregressive manner, the NARRrepair model greatly increases the inference speed without significantly decreasing the number of repaired programs. More specifically, NARRepair fixes 69 bugs from the Defect4J v1.2 dataset, which represents 90\% of the optimal model (Tare); NARRepair fixes 45 bugs from the Defect4J v2.0 dataset, which represents 95\% of the optimal model (Rewardrepair); NARRepair fixes 23 bugs from the QuixBugs dataset, which represents 82\% of the optimal model (AlphaRepair). Moreover, the performance of the NARRrepair model is even better than that of some autoregressive APR models, such as CoCoNut. Overall, this result demonstrates that 1) the NAR model in the field of machine translation cannot be directly used for the APR task, and 2) the proposed NARRepair model is effective for the APR task.

\vspace{1.0mm}
\noindent
\textbf{Results without Perfect Fault Localization.}
We also compare the performance of the NARRepair model with that of the baseline models when the defect location is not known in advance. For this, we use Ochiai \cite{abreu2007accuracy}, a widely used spectrum-based fault localization tool to establish the suspiciousness scores of buggy statements and rank them accordingly. Since the Ochiai tool cannot perfectly locate all bugs, the number of fixed bugs by all AR models and NAR models  will decrease. The detailed results are also shown in Table \ref{tab:t1}. Under this scenario, the NARRrepair model correctly repairs 51, 26, and 20 bugs for the three datasets Defect4J v1.2, Defect4J v2.0, and QuixBugs respectively. Overall, the NARRepair model demonstrates similar results as in the perfect fault localization scenario. First, compared with other NAR models, the performance of the NARRepair model is improved by more than 150\% for the three datasets. Second, compared with DL-based APR techniques that use the autoregressive manner, the NARRepair model delivers comparable performance. 

\begin{figure}
\centering
\includegraphics[width=0.80\linewidth]{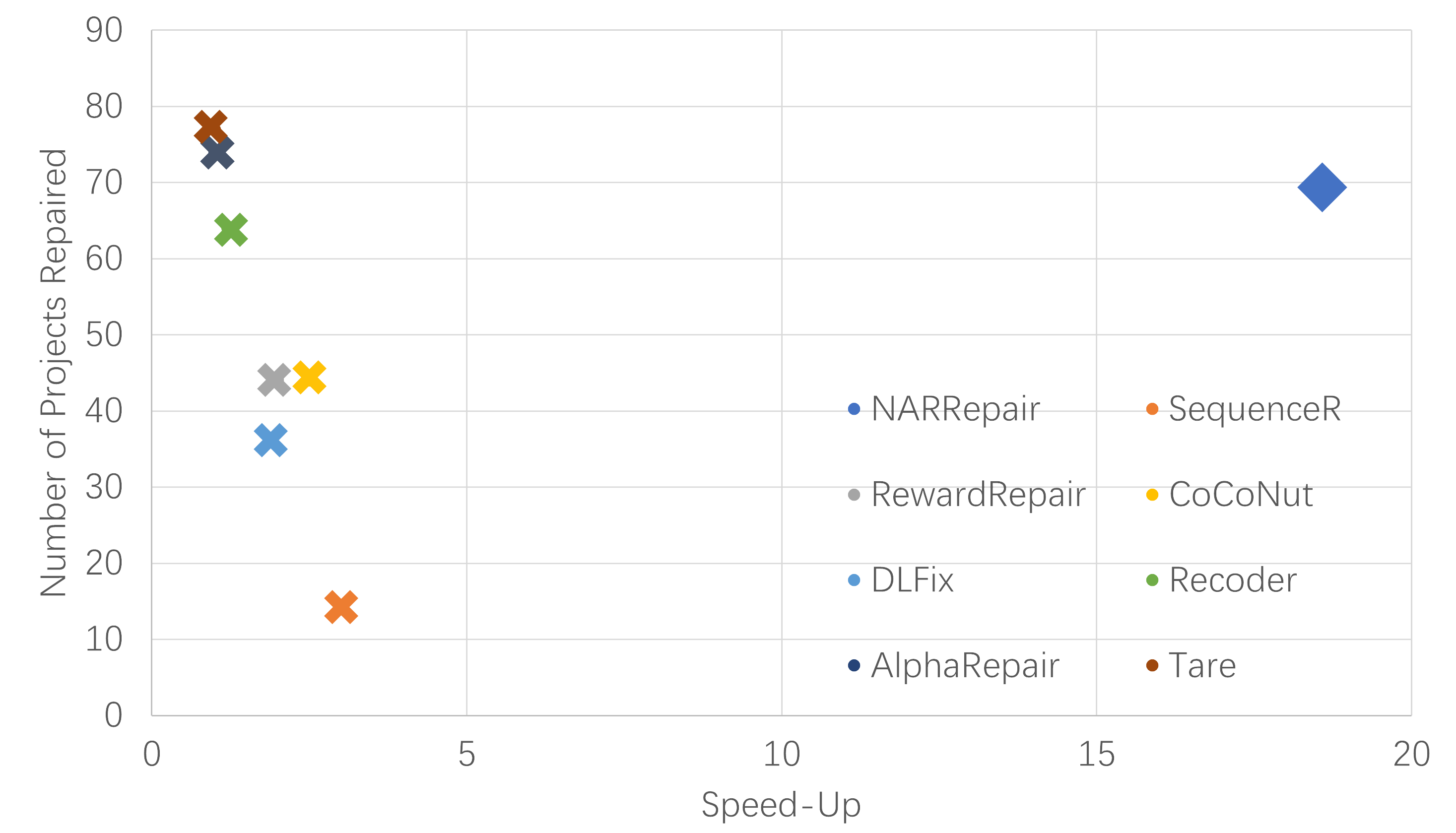}
\caption{\label{fig:frog8}The scatter plot of the number of repaired bugs and the inference speed for different APR models on the Defect v1.2 dataset.}
\end{figure}

\begin{table}[]
\renewcommand{\arraystretch}{1.5}
\caption{\label{tab:t2}Inference time of model with patch evaluation time}
\resizebox{\textwidth}{!}{
\begin{tabular}{ccccccccc}
\hline
Model    & SequenceR        & CoCoNut     & Rewardrepair & Recoder     & AlphaRepair  & Tare         & NARRepair \\ \hline
Latency on TestSet & 1.0x     & -1.2x(5.1x) & -1.4x(5.9x)  & -2.4x(10.1x) & -2.7x(11.3x) & -2.5x(10.5x) & 4.2x      \\ \hline
\end{tabular}}
\end{table}
\begin{table}[]
\renewcommand{\arraystretch}{1.5}
\caption{\label{tab:t3}The parameters of the APR models}

\begin{tabular}{cccccccc}
\hline
Model      & SequenceR & CoCoNut & Rewardrepair & Recoder & AlphaRepair & Tare & NARRepair \\ \hline
Parameters & 4M        & 80M     & 160M         & 54M      & 140M        & 70M  & 150M      \\ \hline
\end{tabular}

\end{table}
\vspace{1.0mm}
\noindent
\textbf{Results of Model Inference Speed.}
We evaluate the inference time of the NARRepair model and the baseline models in both CPU and GPU environments. When calculating the inference time, we let each of the baseline models and the NARRepair model generate 200 patches and calculate the average time to generate each patch. In the experiment, we use the inference time of DL-based APR model SequenceR, which has the fastest inference speed, as the baseline and show the results in Table \ref{tab:t1}. Compared with DL-based APR models with complex structure, such as AlphaRepair, the inference speed of NARRepair is increased by 18.6 times in the GPU environment and 15.2 times in the CPU environment. Compared with DL-based APR models with simple structure, such as SequenceR, the inference speed of NARRepair is still increased by 5.4 and 6.2 times in the CPU and GPU environments respectively. Overall, the results show that compared with other DL-based APR models, the inference speed of the NARRrepair model has been greatly improved. Figure \ref{fig:frog8} displays the scatter plot of the inference speed and the number of repaired bugs for different DL-based APR models and the NARRepair model. We can see that the NARRepair model is located on the top-right of the figure, which suggests that the NARRepair model is superior to other DL-based APR models when both repair accuracy and inference speed are taken into account. With regard to other NAR models, the inference speed of NARRepair is slower in comparison due to the intermediate modules added. However, note that the NARRepair model accounts for both the accuracy and inference speed well and meets the needs of the APR task. 

For the application of the program repair tool in practice, the tool may test the generated patches during the patch generation process and present the patches that pass all tests to the programmers. Considering this, we additionally add the test time required for each patch to the inference time in Table \ref{tab:t1} and recalculate the whole time of the NARRepair model and the baseline models. According to our test evaluation results, there exist no significant differences between the test evaluation time for patches generated by different models. Table \ref{tab:t2} gives the inference time with patch evaluation in detail. Since the test time of each model is the basically the same, the gap in inference speed between the NARRepair model and the baseline models is reduced. However, note that the inference speed of NARRepair is still 4.2 to 11.3 times faster than that of the baseline models. This result again suggests the superiority of the NARRepair model in terms of inference speed.

We also consider the impact of the number of parameters of the model on the inference speed. Generally speaking, the smaller the number of model parameters, the fewer calculations required during the inference process and the faster the inference speed. To verify that the inference speed of the NARRepair model has nothing to do with the number of parameters of the model, we count the number of parameters of all baseline models and show them in Table \ref{tab:t3}. From the data in Table \ref{tab:t3}, we can find that, except for SequenceR, the parameters of other models are at the same level. Overall, the NARRepair model has more parameters than most baseline models. This suggests that the acceleration of the inference speed of the NARRepair model is not due to the reduction in the number of parameters.

\subsection{(RQ2) Results of Ablation Study}
To evaluate the contribution of each part of the NARRepair model, we perform an ablation study on the Defect4J v1.2 dataset under the perfect fault localization scenario. Starting from the complete model, we remove a specific part of the model structure each time and observe the impact of the removal on the results. More specifically, we (1) first replace the two-stage encoder with a normal NAR encoder to observe the impact of contextual information on the results; (2) then replace the attention mechanism for fusing inter-word dependency information with vector addition to observe the role of the attention mechanism; (3) next remove the inter-word dependency extractor to observe the impact of inter-word dependency information on the results; (4) finally remove the repair action predictor and instead pass only the repair length to the decoder to observe the impact of repair actions on the results. The results are shown in Table \ref{tab:t4}.
\begin{table}[]
\renewcommand{\arraystretch}{1.5}
\caption{\label{tab:t4}Ablation Study on Defect4J v1.2 Dataset}
\begin{tabular}{lcc}
\hline
Model                             & \multicolumn{1}{l}{Perfect FL}\\ \hline
NARRepair                & 69                                                               \\
--Two-stage Decoder               & 54                                                               \\
--Attention Layer                 & 49                                                                \\
--Inter-word Dependency Extractor & 40                                                              \\
--Repair Action Predictor         & 34                                                              \\ \hline
\end{tabular}
\end{table}

From the results in the table, we draw the following conclusions. First, after removing the two-stage decoder, the number of repaired programs by the model drops by 15 under the perfect fault localization scenario. This shows that the contextual information obtained by the decoder through the [Mask] tag is of great significance for the result. Second, after removing the feature fusion of the attention mechanism, the performance of the model decreases slightly for the evaluation scenario. Third, after removing the inter-word dependency extractor, the number of repaired programs by the model drops by 9 under the perfect fault localization scenario. This result verifies that the obtained inter-word dependency information can effectively improve the prediction accuracy of the model. Finally, after removing the repair action predictor, the performance of the model drops significantly. Judging from the experimental results, among all modules of the model, the repair action predictor has the greatest impact on the model performance. This implies that the prediction of repair actions can effectively avoid the problem of modifying correct words into wrong ones. It also suggests that this is one of the main reasons why the naive NAR model performs poorly for the APR task.

\subsection{(RQ3) Results of Loading Pre-trained Model Weights}
To show that our method can help pre-trained models improve inference speed, we select some mainstream pre-trained models for programming languages, including CodeBERT\cite{feng2020codebert}, GraphCodeBERT\cite{guo2020graphcodebert}, CodeT5\cite{wang2021codet5}, and CodeGPT\cite{lu2021codexglue}. Among them, CodeBERT and GraphCodeBERT are composed of encoders of Transformers, CodeT5 is composed of both encoders and decoders of Transformers, and CodeGPT is composed of decoders of Transformers. Since NARRepair is also based on the Transformer structure, we can respectively load the weights of the four pre-trained models into the NARRepair model. To observe the impact of our method on the pre-trained model, we compare the results generated using the pre-trained model with those generated by NARRepair, which has loaded the weights of the pre-trained model. We conduct these experiments on the Defect4J v1.2 dataset under the perfect fault localization scenario, and the results are shown in Table \ref{tab:t5}.
\begin{table}[]
\renewcommand{\arraystretch}{1.5}
\caption{\label{tab:t5}The Results of Loading Pre-trained Model Weights}
\resizebox{0.7\textwidth}{!}{
\begin{tabular}{lccc}
\hline
Model                   & \multicolumn{1}{l}{Correct Project} & \multicolumn{1}{l}{performance ratio} & \multicolumn{1}{l}{Speed-Up} \\ \hline
CodeBERT                & 43                                  & \multirow{2}{*}{90.7\%}                & 1x                                     \\
NARRepair+CodeBERT      & 39                                  &                                       & 6.3x                                   \\ \hline
GraphCodeBERT           & 54                                  & \multirow{2}{*}{92.6\%}                & 1x                                     \\
NARRepair+GraphCodeBERT & 50                                  &                                       & 6.3x                                   \\ \hline
CodeT5                  & 73                                  & \multirow{2}{*}{94.5\%}                & 1x                                     \\
NARRepair+CodeT5        & 69                                  &                                       & 6.4x                                   \\ \hline
CodeGPT                 & 49                                  & \multirow{2}{*}{85.7\%}                & 1x                                     \\
NARRepair+CodeGPT       & 42                                  &                                       & 6.2x                                   \\ \hline
\end{tabular}}
\end{table}

According to the data in the table, we find that compared with the four pre-trained models, the inference speed of the NARRepair model is 6.2 times to 6.4 times faster. After loading the weights of the four pre-trained models, the NARRepair model can still achieve 85.7\% to 94.5\% of their performance obtained using the AR manner. This result suggests that the NARRepair model can help the pre-trained models significantly improve their inference speed without significantly reducing performance. The large language model (\emph{e.g.}, \cite{touvron2023llama}) is very similar to the pre-trained model except for the number of parameters. However, due to the large number of parameters of the large language models and device limitations, we cannot load the parameters of the large language models into NARRepair. Through the results for pre-trained models, however, we can to certain degree deduce that NARRepair can help large language models significantly improve inference speed. 

 \subsection{(RQ4) Results of Alleviating the Over-Correction Problem}
In Section 3, we mentioned that one of the main purposes of the repair action predictor and the two-stage decoder is to avoid changing correct words into wrong ones during the repair process. The repair action predictor avoids modifying correct words by predicting that the repair action for those words is ``Keep''(§3.2). The two-stage decoder decodes words with low confidence again to avoid some correct words from being modified (§3.4). To explore the effectiveness of our idea in more detail, we respectively remove the repair action predictor module and the two-stage decoder module, and observe the corresponding changes in the output results of the NARRepair model. Among all patches generated by the NARRepair model on the Defect4j v1.2 dataset, we count the average number of correct words changed into incorrect ones and present the results in Table \ref{tab:t6}.
\begin{table}[]
\renewcommand{\arraystretch}{1.5}
\caption{\label{tab:t6}The number of correct words modified}
\setlength{\tabcolsep}{10mm}{
\begin{tabular}{ll}
\hline
Model                     & Count \\ \hline
NARRepair                 & 3.1   \\ 
--Repair Action Predictor & 3.9   \\ 
--Two-stage Decoder       & 6.2   \\ \hline
\end{tabular}
}
\end{table}

From the results in Table \ref{tab:t6}, we can observe that both the repair action predictor module and the two-stage decoder module reduce the number of correct words in the code that are modified into incorrect ones. When we remove the repair action predictor from the NARRepair model, the number of correct words modified in the patches (generated by the model) increases by 0.8 on average. When we remove the two-stage decoder at the same time, the number of correct words modified in the patches (generated by the NARRepair model) further increases by 2.3 on average. This suggests that the two-stage decoder is more effective in avoiding correct words wrongly modified into incorrect ones. However, this does not imply that the repair action predictor plays an unimportant role in the entire model. Through the ablation study, we have shown the importance of the repair action predictor. Overall, the experimental result suggests that NARRepair can effectively prevent correct words from being modified during the model repair process.

\begin{figure}
\centering
\includegraphics[width=0.80\linewidth]{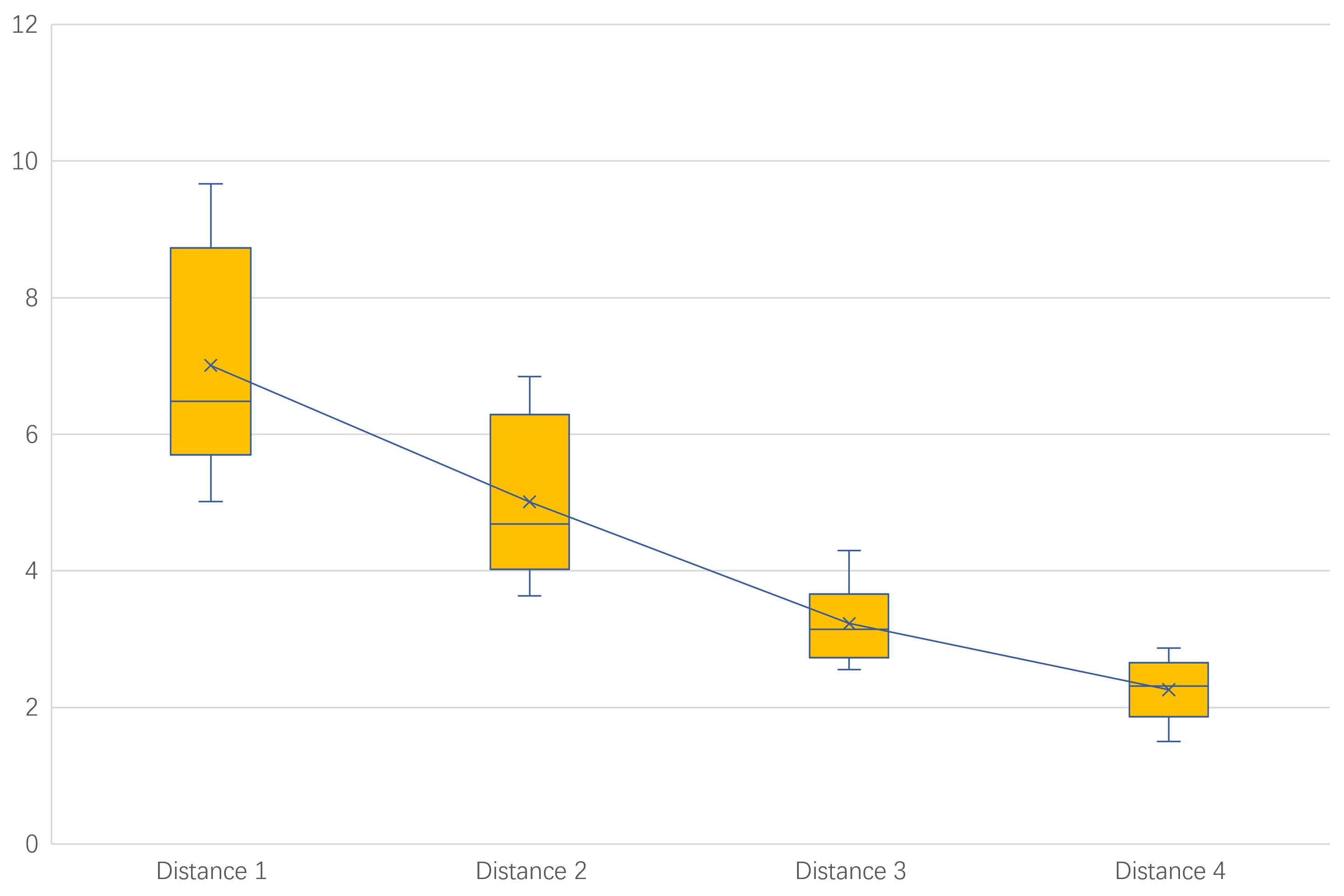}
\caption{\label{fig:frog9}The cosine similarity between nodes in the abstract syntax tree under different distance values. }
\end{figure}
\subsection{(RQ5) Results of the Similarity Between AST Nodes}
To show the degree of correlation between parent nodes and child nodes in the abstract syntax tree, we introduce cosine similarity. Usually, when the cosine similarity of two feature vectors is high, it means that the two vectors are strongly related, and vice versa. We obtain the inter-word dependency matrix output by the trained NARRepair model. Each content in the dependency matrix represents a parent node in the AST. We calculate the feature cosine similarity between the feature vector of the parent node in the dependency matrix and that of each word in the code text (generated by the inter-word dependency extractor). We classify the calculated values by the distance from parent nodes to child nodes in the AST, and the results are shown in Figure \ref{fig:frog9}. According to Figure \ref{fig:frog9}, we can see that the similarity between the parent node and the child node gradually decreases as the distance increases. We also find that the farther the distance between the parent node and the child node, the greater the fluctuation range of similarity. 

\begin{figure}
\centering
\includegraphics[width=0.80\linewidth]{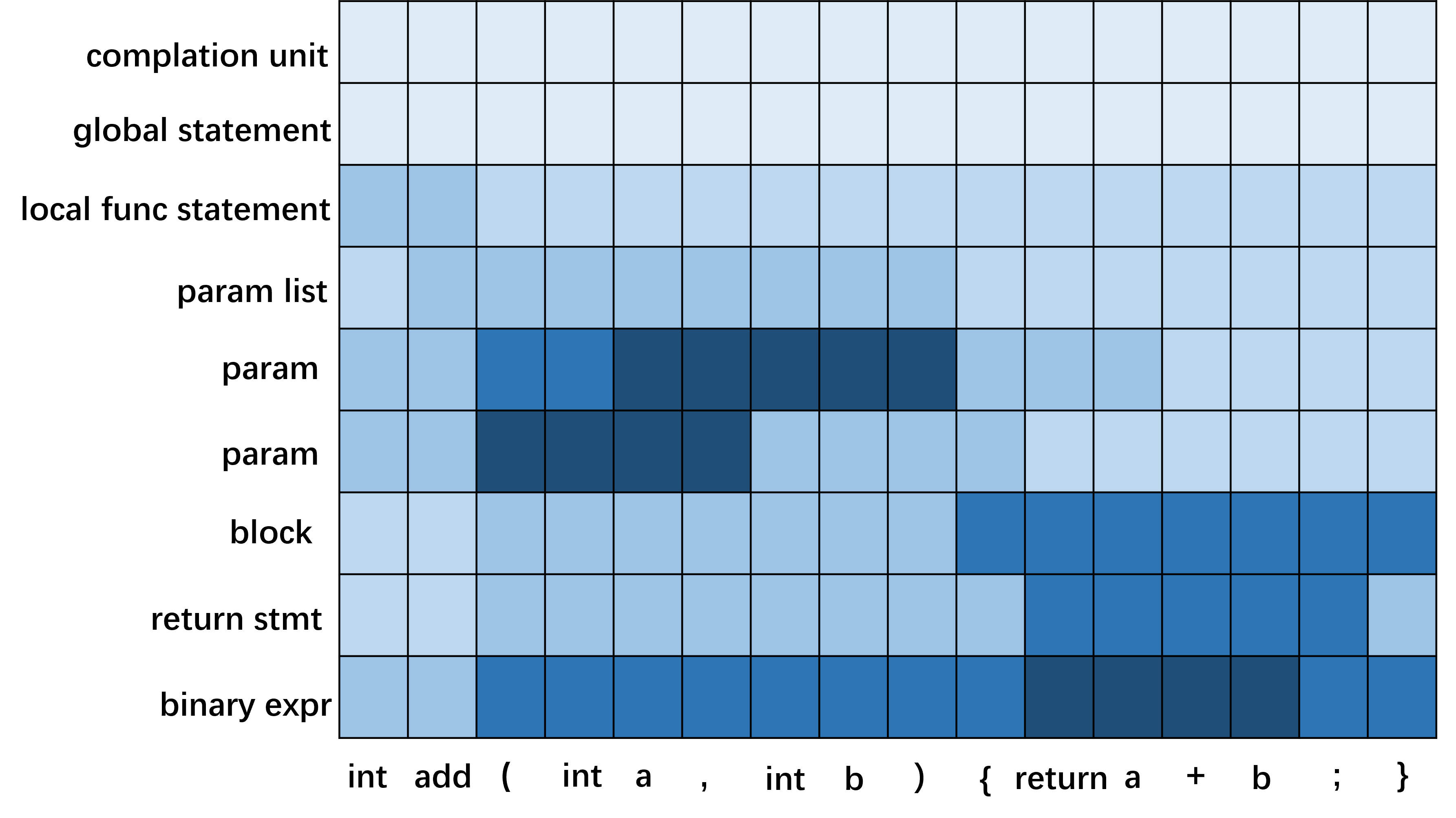}
\caption{\label{fig:frog10}Heat map of the cosine similarity between nodes in the AST of the code text ``int add (int a, int b) \{return a+b;\}''}
\end{figure}

To show the relationship between parent nodes and child nodes more concretely, we take the code text ``int add (int a, int b) \{return a+b;\}'' as an example and calculate the cosine similarity between the parent node and the child nodes in its AST. The calculated results are shown in Figure \ref{fig:frog10} as a heat map. In the heat map, the darker the color of the square, the higher the cosine similarity between the vectors and the closer the relationship between the two vectors. From Figure \ref{fig:frog10}, we can see that the cosine similarities between the parent node ``binary expr'' and its child nodes ``a'', ``+'', and ``b'' are all high, and the similarity between it and other words decreases as the distance increases. The nodes ``int a'' and ``int b'' and their parent node ``param'' also comply with this trend. These results verify our hypothesis that the relationship between parent nodes and child nodes is close and the closeness decreases with the increase of distance. In addition, this result also justifies the rationality of our method of using the nearest common parent node as the dependency relationship between words.

\section{DISCUSSION}

\subsection{Threats to Validity}
In line with existing works on automatic program repair, our results should be interpreted with several threats to validity in mind.

\textbf{Internal Validity.} Threats to internal validity might come from the potential faults in the implementations of NARRepair itself and its evaluation. To avoid faults in the implementation of NARRepair, our implementation is mainly based on mature deep learning and program analysis libraries, such as 
Fairseq \cite{ott-etal-2019-fairseq} and Tree-Sitter \cite{tree-sitter}. In addition, we have performed thorough testing to ensure the correctness of NARRepair. To alleviate the threats to evaluation, we used the reported results published by the respective papers for the baselines to facilitate fair comparison. When running the models for the baselines is necessary, we re-run the models provided by the authors. Furthermore, note that the whole artifact related with this article is made available online for scrutiny and extension by other researchers. 

\textbf{External Validity.} A potential threat to external validity concerns the representativeness of the benchmark used in our experiment. To mitigate this threat as much as possible, 
we used three widely used datasets in the APR community for evaluating NARRepair, including Defects4J v1.2, Defects4J v2.0, and QuixBugs. The obtained results illustrate the effectiveness and generalizability of NARRepair. However, we believe that additional evaluations on other benchmarks (such as \cite{9401985}) can further confirm its effectiveness and generalizability. In addition, since the used three datasets contain only Java bugs, further studies are needed to investigate the effectiveness of NARRepair on other programming languages.


\subsection{Future Work}
For future work, we will focus particularly on achieving higher accuracy while increasing the inference speed, even better than that of the state-of-the-art AR models. In addition, we will evaluate the performance of NARRepair using more benchmarks and additional programming languages to verify the effectiveness and generalizability of the model. Finally, we also plan to apply our key ideas to other software engineering tasks for which the inference speed is of great importance. Indeed, we hope that our work can arouse the interest of software engineering researchers in the inference speed of deep learning models for software engineering tasks, which can make software engineering research better meet actual developer needs.

\section{Conclusions}
In this paper, to increase the inference speed while maintaining the accuracy of repairing buggy code text, we propose NARRepair, a non-autoregressive model for automatic program repair. To solve the issues of wrongly modifying correct words into wrong ones, missing inter-word dependency information, and missing contextual information that are generally associated with non-autoregressive models, we propose a repair action predictor, inter-word dependency extractor, and two-stage decoder in NARRepair for addressing the three issues respectively. We evaluate the performance of the NARRepair model on three widely used datasets for automatic program repair tasks, including Defect4J v1.2, Defect4J v2.0, and QuixBugs. The experimental results show that while maintaining the high repair accuracy, NARRepair's inference speed is 5.4 to 18.6 times faster than other automatic program repair models. Besides, the results additionally show that NARRepair performs better than other non-autoregressive models. Finally, we also demonstrate that our method is effective for pre-trained models and is suitable for generalizing to large language models in order to meet the need for real-time code repair.

\begin{acks}
\end{acks}

\bibliographystyle{ACM-Reference-Format}
\bibliography{sample-base}

\appendix

\end{document}